\pdfoutput=1
\documentclass[12pt,a4paper]{article}
\usepackage{ifthen} % for conditional statements
\newboolean{pdflatex}
\setboolean{pdflatex}{true} % False for eps figures 

\newboolean{articletitles}
\setboolean{articletitles}{true} % False removes titles in references

\newboolean{uprightparticles}
\setboolean{uprightparticles}{false} %True for upright particle symbols

\newboolean{inbibliography}
\setboolean{inbibliography}{false} %True once you enter the bibliography

\usepackage[top=1in, bottom=1.25in, left=1in, right=1in]{geometry}

\usepackage{microtype}
\usepackage{lineno}  % for line numbering during review
\usepackage{xspace} % To avoid problems with missing or double spaces after
\usepackage{caption} %these three command get the figure and table captions automatically small

\usepackage{graphicx}  % to include figures (can also use other packages)
\usepackage{color}
\usepackage{colortbl}
\graphicspath{{./figs/}} % Make Latex search fig subdir for figures

\usepackage{amsmath} % Adds a large collection of math symbols
\usepackage{amssymb}
\usepackage{amsfonts}
\usepackage{upgreek} % Adds in support for greek letters in roman typeset

\newcommand*\patchAmsMathEnvironmentForLineno[1]{%
\expandafter\let\csname old#1\expandafter\endcsname\csname #1\endcsname
\expandafter\let\csname oldend#1\expandafter\endcsname\csname
end#1\endcsname
 \renewenvironment{#1}%
   {\linenomath\csname old#1\endcsname}%
   {\csname oldend#1\endcsname\endlinenomath}%
}
\newcommand*\patchBothAmsMathEnvironmentsForLineno[1]{%
  \patchAmsMathEnvironmentForLineno{#1}%
  \patchAmsMathEnvironmentForLineno{#1*}%
}
\AtBeginDocument{%
\patchBothAmsMathEnvironmentsForLineno{equation}%
\patchBothAmsMathEnvironmentsForLineno{align}%
\patchBothAmsMathEnvironmentsForLineno{flalign}%
\patchBothAmsMathEnvironmentsForLineno{alignat}%
\patchBothAmsMathEnvironmentsForLineno{gather}%
\patchBothAmsMathEnvironmentsForLineno{multline}%
\patchBothAmsMathEnvironmentsForLineno{eqnarray}%
}

\usepackage{hyperref}    % Hyperlinks in references
\usepackage[all]{hypcap} % Internal hyperlinks to floats.

\usepackage{xspace} 
\usepackage{upgreek}

\def\lhcb {\mbox{LHCb}\xspace}

\def\belle  {\mbox{Belle}\xspace}

\def\dzero  {\mbox{D0}\xspace}

\def\MagUp {\mbox{\em Mag\kern -0.05em Up}\xspace}

\ifthenelse{\boolean{uprightparticles}}%
{

 \def\Ppi         {\ensuremath{\uppi}\xspace}

 \def\PDelta      {\ensuremath{\Delta}\xspace}                 
 \def\PXi      {\ensuremath{\Xi}\xspace}                 
 \def\PLambda      {\ensuremath{\Lambda}\xspace}                 
 \def\PSigma      {\ensuremath{\Sigma}\xspace}                 
 \def\POmega      {\ensuremath{\Omega}\xspace}                 
 \def\PUpsilon      {\ensuremath{\Upsilon}\xspace}                 
 
 \def\PB      {\ensuremath{\mathrm{B}}\xspace}                 
                  
 \def\PD      {\ensuremath{\mathrm{D}}\xspace}

 \def\PK      {\ensuremath{\mathrm{K}}\xspace}

 \def\Pb      {\ensuremath{\mathrm{b}}\xspace}                 
 \def\Pc      {\ensuremath{\mathrm{c}}\xspace}

 \def\Pi      {\ensuremath{\mathrm{i}}\xspace}

 \def\Ps      {\ensuremath{\mathrm{s}}\xspace}

}
{

 \def\Ppi         {\ensuremath{\pi}\xspace}

 \mathchardef\PDelta="7101
 \mathchardef\PXi="7104
 \mathchardef\PLambda="7103
 \mathchardef\PSigma="7106
 \mathchardef\POmega="710A
 \mathchardef\PUpsilon="7107
                  
 \def\PB      {\ensuremath{B}\xspace}                 
                  
 \def\PD      {\ensuremath{D}\xspace}

 \def\PK      {\ensuremath{K}\xspace}

 \def\Pb      {\ensuremath{b}\xspace}                 
 \def\Pc      {\ensuremath{c}\xspace}

 \def\Pi      {\ensuremath{i}\xspace}

 \def\Ps      {\ensuremath{s}\xspace}

}

\makeatletter
\ifcase \@ptsize \relax% 10pt
  \newcommand{\miniscule}{\@setfontsize\miniscule{4}{5}}% \tiny: 5/6
\or% 11pt
  \newcommand{\miniscule}{\@setfontsize\miniscule{5}{6}}% \tiny: 6/7
\or% 12pt
  \newcommand{\miniscule}{\@setfontsize\miniscule{5}{6}}% \tiny: 6/7
\fi
\makeatother

\DeclareRobustCommand{\optbar}[1]{\shortstack{{\miniscule (\rule[.5ex]{1.25em}{.18mm})}
  \\ [-.7ex] $#1$}}

\def\squark    {{\ensuremath{\Ps}}\xspace}

\def\cquark    {{\ensuremath{\Pc}}\xspace}

\def\bquark    {{\ensuremath{\Pb}}\xspace}

\def\pion   {{\ensuremath{\Ppi}}\xspace}

\def\pip    {{\ensuremath{\pion^+}}\xspace}
\def\pim    {{\ensuremath{\pion^-}}\xspace}
\def\pipm   {{\ensuremath{\pion^\pm}}\xspace}
\def\pimp   {{\ensuremath{\pion^\mp}}\xspace}

\def\kaon    {{\ensuremath{\PK}}\xspace}
  \def\Kbar    {{\kern 0.2em\overline{\kern -0.2em \PK}{}}\xspace}

\def\KorKbar    {\kern 0.18em\optbar{\kern -0.18em K}{}\xspace}
\def\Kz      {{\ensuremath{\kaon^0}}\xspace}
\def\Kzb     {{\ensuremath{\Kbar{}^0}}\xspace}
\def\Kp      {{\ensuremath{\kaon^+}}\xspace}
\def\Km      {{\ensuremath{\kaon^-}}\xspace}
\def\Kpm     {{\ensuremath{\kaon^\pm}}\xspace}

\def\KS      {{\ensuremath{\kaon^0_{\mathrm{ \scriptscriptstyle S}}}}\xspace}

  \def\Dbar    {{\kern 0.2em\overline{\kern -0.2em \PD}{}}\xspace}
\def\D       {{\ensuremath{\PD}}\xspace}

\def\DorDbar    {\kern 0.18em\optbar{\kern -0.18em D}{}\xspace}
\def\Dz      {{\ensuremath{\D^0}}\xspace}

\def\Dpm     {{\ensuremath{\D^\pm}}\xspace}

\def\Dspm    {{\ensuremath{\D^{\pm}_\squark}}\xspace}

\def\B       {{\ensuremath{\PB}}\xspace}
\def\Bbar    {{\ensuremath{\kern 0.18em\overline{\kern -0.18em \PB}{}}}\xspace}

\def\BorBbar    {\kern 0.18em\optbar{\kern -0.18em B}{}\xspace}

\def\Bpm     {{\ensuremath{\B^\pm}}\xspace}

  \def\Y#1S{\ensuremath{\PUpsilon{(#1S)}}\xspace}% no space before {...}!

\def\Lz          {{\ensuremath{\PLambda}}\xspace}
\def\Lbar        {{\ensuremath{\kern 0.1em\overline{\kern -0.1em\PLambda}}}\xspace}
\def\LorLbar    {\kern 0.18em\optbar{\kern -0.18em \PLambda}{}\xspace}

\def\Lb      {{\ensuremath{\Lz^0_\bquark}}\xspace}

\def\BF         {{\ensuremath{\mathcal{B}}}\xspace}

\def\BR         {\BF}
\def\to                 {\ensuremath{\rightarrow}\xspace}

\def\CP                {{\ensuremath{C\!P}}\xspace}

\newcommand{\ACP}{{\ensuremath{{\mathcal{A}}^{\CP}}}\xspace}

\def\AT#1     {\ensuremath{A_{\mathrm{T}}^{#1}}\xspace}           % 2

\def\C#1      {\ensuremath{\mathcal{C}_{#1}}\xspace}                       % 9
\def\Cp#1     {\ensuremath{\mathcal{C}_{#1}^{'}}\xspace}                    % 7
\def\Ceff#1   {\ensuremath{\mathcal{C}_{#1}^{\mathrm{(eff)}}}\xspace}        % 9  
\def\Cpeff#1  {\ensuremath{\mathcal{C}_{#1}^{'\mathrm{(eff)}}}\xspace}       % 7
\def\Ope#1    {\ensuremath{\mathcal{O}_{#1}}\xspace}                       % 2
\def\Opep#1   {\ensuremath{\mathcal{O}_{#1}^{'}}\xspace}                    % 7

\newcommand{\tev}{\ifthenelse{\boolean{inbibliography}}{\ensuremath{~T\kern -0.05em eV}\xspace}{\ensuremath{\mathrm{\,Te\kern -0.1em V}}}\xspace}
\newcommand{\gev}{\ensuremath{\mathrm{\,Ge\kern -0.1em V}}\xspace}
\newcommand{\mev}{\ensuremath{\mathrm{\,Me\kern -0.1em V}}\xspace}
\newcommand{\kev}{\ensuremath{\mathrm{\,ke\kern -0.1em V}}\xspace}
\newcommand{\ev}{\ensuremath{\mathrm{\,e\kern -0.1em V}}\xspace}
\newcommand{\gevc}{\ensuremath{{\mathrm{\,Ge\kern -0.1em V\!/}c}}\xspace}
\newcommand{\mevc}{\ensuremath{{\mathrm{\,Me\kern -0.1em V\!/}c}}\xspace}
\newcommand{\gevcc}{\ensuremath{{\mathrm{\,Ge\kern -0.1em V\!/}c^2}}\xspace}
\newcommand{\gevgevcccc}{\ensuremath{{\mathrm{\,Ge\kern -0.1em V^2\!/}c^4}}\xspace}
\newcommand{\mevcc}{\ensuremath{{\mathrm{\,Me\kern -0.1em V\!/}c^2}}\xspace}

\def\mum  {\ensuremath{{\,\upmu\mathrm{m}}}\xspace}

\def\invfb   {\ensuremath{\mbox{\,fb}^{-1}}\xspace}

\def\gsim{{~\raise.15em\hbox{$>$}\kern-.85em
          \lower.35em\hbox{$\sim$}~}\xspace}
\def\lsim{{~\raise.15em\hbox{$<$}\kern-.85em
          \lower.35em\hbox{$\sim$}~}\xspace}

\def\ptot       {\mbox{$p$}\xspace}
\def\pt         {\mbox{$p_{\mathrm{ T}}$}\xspace}

\def\evtgen     {\mbox{\textsc{EvtGen}}\xspace}

\def\geant      {\mbox{\textsc{Geant4}}\xspace}

\def\photos     {\mbox{\textsc{Photos}}\xspace}

\def\pythia     {\mbox{\textsc{Pythia}}\xspace}

\def\tell1  {TELL1\xspace}
\def\ukl1   {UKL1\xspace}

\usepackage{xspace} 
\usepackage{upgreek}

\newcommand{\Dcand}{\ensuremath{D^{\pm}_{(s)}}}
\newcommand{\DcandP}{\ensuremath{D^{+}_{(s)}}}
\newcommand{\DcandM}{\ensuremath{D^{-}_{(s)}}}

\newcommand{\DcandToEtapPi}{\ensuremath{\Dcand\rightarrow\Etap\pipm}} 
\newcommand{\DToEtapPi}{\ensuremath{\Dpm\rightarrow\Etap\pipm}} 
\newcommand{\DsToEtapPi}{\ensuremath{\Dspm\rightarrow\Etap\pipm}} 

\newcommand{\DToKsPi}{\ensuremath{\Dpm\rightarrow\KS\pipm}}

\newcommand{\DsToPhiPi}{\ensuremath{\Dspm\rightarrow\phi\pipm }}

\newcommand{\EtapPi}{\ensuremath{\eta^\prime\pi}}
\newcommand{\Etap}{\ensuremath{\eta^\prime}}
\newcommand{\EtapPPG}{\ensuremath{\Etap\rightarrow\pip\pim\gamma}}

\newcommand{\DandDsToPhiPi}{\ensuremath{\Dcand\rightarrow\phi\pipm}}

\newcommand{\DToPhiPiPPP}{\ensuremath{\Dcand\rightarrow\phi_{3\pi}\pipm }}

\newcommand{\DpmToPhiPiPPP}{\ensuremath{\Dpm\rightarrow\phi_{3\pi}\pipm }}
\newcommand{\DspmToPhiPiPPP}{\ensuremath{\Dspm\rightarrow\phi_{3\pi}\pipm }}

\newcommand{\DcandToPiPiPi}{\ensuremath{\Dcand\rightarrow\pimp\pipm\pipm }}

\newcommand{\DzToKpKm}{\ensuremath{\Dz\rightarrow\Kp\Km}}
\newcommand{\DzToPipPim}{\ensuremath{\Dz\rightarrow\pip\pim}}

\renewcommand{\ACP}{\ensuremath{\mathcal{A}_{\CP}}}
\newcommand{\DeltaACP}{\ensuremath{\Delta \mathcal{A}_{\CP}}}

\newcommand{\ACPprod}{\mathcal{A}_{\mathrm{P}}}
\newcommand{\ACPdet}{\mathcal{A}_{\mathrm{D}}}
\newcommand{\ACPraw}{\ensuremath{\mathcal{A}_{\mathrm{raw}}}}

\newcommand{\ACPKzKzb}{\mathcal{A}({\Kzb-\Kz})}

\usepackage{mciteplus}

\usepackage{longtable} % only for template; not usually to be used in PAPERs

\begin{document}

%%%%%%%%%%%%%%%%%%%%%%%%%
%%%%% Title     %%%%%%%%%
%%%%%%%%%%%%%%%%%%%%%%%%%
\renewcommand{\thefootnote}{\fnsymbol{footnote}}
\setcounter{footnote}{1}

% %%%%%%% CHOOSE TITLE PAGE--------
%\onecolumn
%\input{title-LHCb-INT}
%\input{title-LHCb-ANA}
%\input{title-LHCb-CONF}
% $Id: title-LHCb-PAPER.tex 95682 2016-07-21 12:13:58Z michaelt $
% ===============================================================================
% Purpose: LHCb-PAPER journal paper title page template
% Author: 
% Created on: 2010-09-25
% ===============================================================================

%%%%%%%%%%%%%%%%%%%%%%%%%
%%%%%  TITLE PAGE  %%%%%%
%%%%%%%%%%%%%%%%%%%%%%%%%
\begin{titlepage}
\pagenumbering{roman}

% Header ---------------------------------------------------
\vspace*{-1.5cm}
\centerline{\large EUROPEAN ORGANIZATION FOR NUCLEAR RESEARCH (CERN)}
\vspace*{1.5cm}
\noindent
\begin{tabular*}{\linewidth}{lc@{\extracolsep{\fill}}r@{\extracolsep{0pt}}}
\ifthenelse{\boolean{pdflatex}}% Logo format choice
{\vspace*{-2.7cm}\mbox{\!\!\!\includegraphics[width=.14\textwidth]{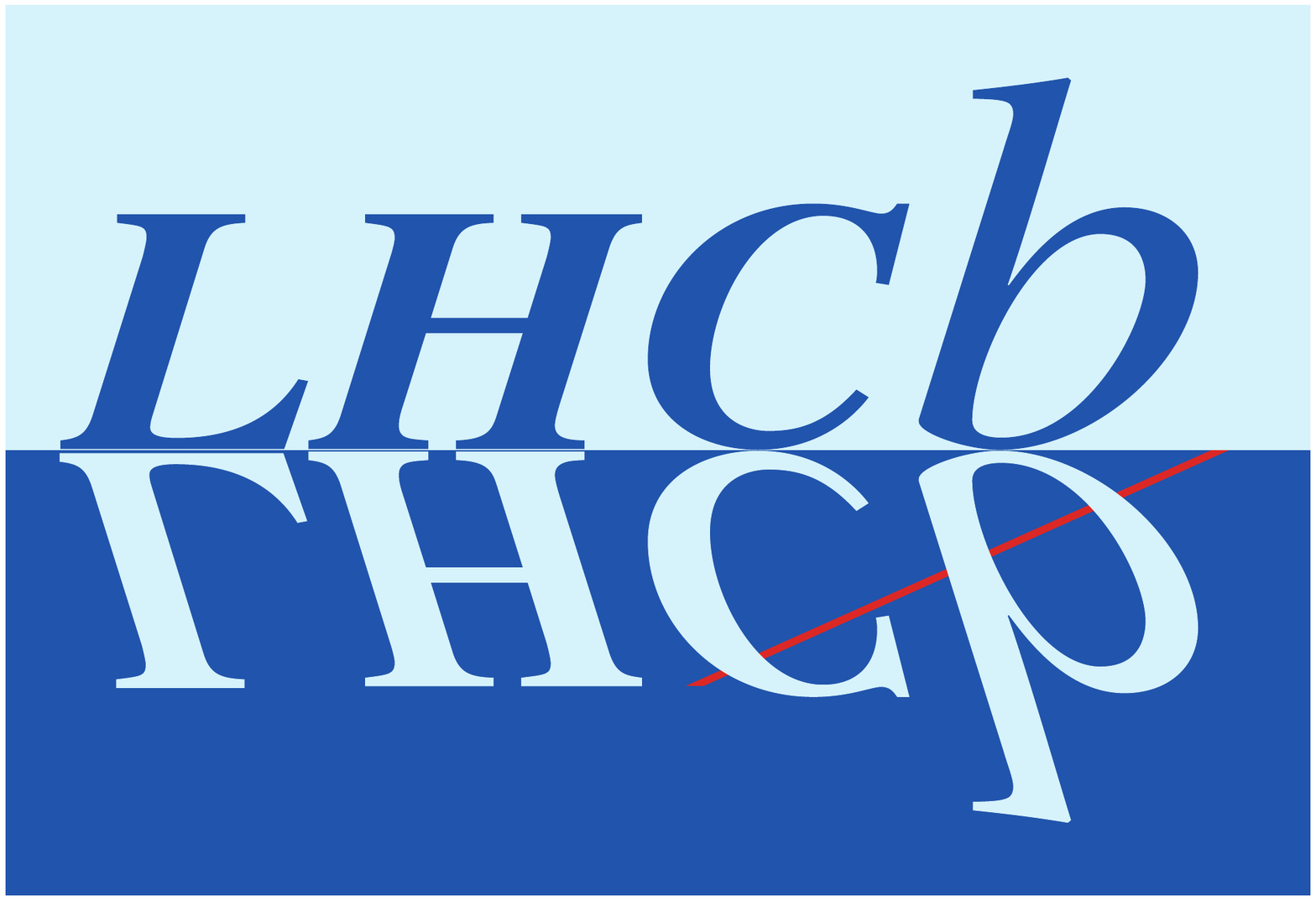}} & &}%
{\vspace*{-1.2cm}\mbox{\!\!\!\includegraphics[width=.12\textwidth]{lhcb-logo.eps}} & &}%
\\
 & & CERN-EP-2016-315 \\  % ID 
 & & LHCb-PAPER-2016-041 \\  % ID 
 & & 2 April 2017 \\ % Date - Can also hardwire e.g.: 23 March 2010
\end{tabular*}

\vspace*{4.0cm}

% Title --------------------------------------------------
{\normalfont\bfseries\boldmath\huge
\begin{center}
  Measurement of $\boldmath{C\!P}$ asymmetries in
  $\boldmath{D^{\pm}\rightarrow \eta^{\prime} \pi^{\pm}}$ and 
  $\boldmath{D_s^{\pm}\rightarrow \eta^{\prime} \pi^{\pm}}$ decays
\end{center}
}

\vspace*{2.0cm}

% Authors -------------------------------------------------
\begin{center}
The LHCb collaboration\footnote{Authors are listed at the end of this Letter.}
\end{center}

\vspace{\fill}

% Abstract -----------------------------------------------
\begin{abstract}  
  \noindent
  A search for \CP\ violation in 
  $D^{\pm}\rightarrow \eta' \pi^{\pm}$ and 
  $D^{\pm}_{s}\rightarrow \eta' \pi^{\pm}$ decays 
  is performed using proton-proton collision data, 
  corresponding to an integrated luminosity of 3\invfb,
  recorded by the LHCb experiment at centre-of-mass energies of $7$~and~$8\,\tev$. 
  The measured \CP-violating 
  charge asymmetries are
  $\ACP(\DToEtapPi)=(-0.61\pm 0.72 \pm 0.53 \pm 0.12)\%$~and $\ACP(\DsToEtapPi)=(-0.82\pm 0.36 \pm 0.22 \pm 0.27)\%$, where 
  the first uncertainties are statistical, the second systematic, and the third 
  are the uncertainties on the 
  $\ACP(\DToKsPi)$ and $\ACP(\DsToPhiPi)$ measurements used for calibration. 
  The results represent the most precise measurements of these 
  asymmetries to date. 

\end{abstract}

\vspace*{2.0cm}

\begin{center}
  Published in Phys.~Lett.~B 771 (2017) 21-30
\end{center}

\vspace{\fill}

{\footnotesize 
\centerline{\copyright~CERN on behalf of the \lhcb collaboration, licence \href{http://creativecommons.org/licenses/by/4.0/}{CC-BY-4.0}.}}
\vspace*{2mm}

\end{titlepage}

%%%%%%%%%%%%%%%%%%%%%%%%%%%%%%%%
%%%%%  EOD OF TITLE PAGE  %%%%%%
%%%%%%%%%%%%%%%%%%%%%%%%%%%%%%%%

%  empty page follows the title page ----
\newpage
\setcounter{page}{2}
\mbox{~}

\cleardoublepage

%\input{title-PLB-PAPER}
%\twocolumn
% %%%%%%%%%%%%% ---------

\renewcommand{\thefootnote}{\arabic{footnote}}
\setcounter{footnote}{0}

%%%%%%%%%%%%%%%%%%%%%%%%%
%%%%% Main text %%%%%%%%%
%%%%%%%%%%%%%%%%%%%%%%%%%

\pagestyle{plain} % restore page numbers for the main text
\setcounter{page}{1}
\pagenumbering{arabic}

%% Uncomment during review phase. 
%% Comment before a final submission.
%\linenumbers

\section{Introduction}
\label{sec:Introduction}

The decays of charmed mesons offer a unique opportunity for the experimental 
investigation of hitherto unobserved \CP\ violation in the up-type quark sector. 
The Standard Model~(SM) predicts \CP\ violation to occur in the charm sector, albeit 
at a level of \ensuremath{\mathcal{O}\left( 10^{-3} \right) } 
at leading order in $1/m_c$, compatible with the lack of evidence in current measurements.
Larger values are possible if new sources of \CP\ violation beyond
the SM exist. The study of charm systems is a unique tool to probe sources of \CP\ violation 
that affect only the dynamics of up-type quarks~\cite{Grossman:2006jg}. 

In order for non-zero \CP\ asymmetries to be observable in a process, 
two or more interfering amplitudes with different \CP-odd and \CP-even phases are needed.
In the SM, no direct \CP\ violation can therefore emerge at leading order in Cabibbo-favoured charm decays, 
which are mediated by a single weak amplitude, while small \CP\ asymmetries are expected 
in singly-Cabibbo-suppressed decays~\cite{Buccella:1996uy} 
due to the interference of colour-allowed tree-level amplitudes with 
loop- (penguin) and colour-suppressed tree-level amplitudes. 
Since these asymmetries may be enhanced by nonperturbative effects~\cite{Golden:1989qx}, 
theoretical interpretations of experimental results 
require the analysis of several channels with similar sensitivity. 
In particular, the study of charm decays to pseudoscalar mesons 
tests flavour topology~\cite{Chau:1982da} and SU(3) predictions, 
and may constrain amplitudes through triangle relations or 
shed light on sources of SU(3) flavour symmetry breaking~\cite{Hinchliffe:1995hz,Bhattacharya:2009ps,Cheng:2012xb}. 
To date, the most precise measurements of \CP\ asymmetries in singly-Cabibbo-suppressed two-body charm decays 
have been performed in \DzToKpKm\ and \DzToPipPim\ decays by the
LHCb collaboration~\cite{LHCb-PAPER-2015-055,LHCB-PAPER-2014-013}, and have shown no evidence for \CP\ violation. 
Among the other charm decays to two pseudoscalar mesons with significant branching fractions, 
thus far \DToEtapPi\ and \DsToEtapPi\ have been studied only 
in $e^+e^-$ collisions~\cite{Onyisi:2013bjt, Won:2011ku} 
due to the experimental difficulty of reconstructing $\eta^{(\prime)}$ mesons in hadron collisions. 
The most recent studies of these decays at the Belle and CLEO experiments 
yielded a \CP asymmetry of $(-0.12\pm 1.12\pm 0.17)\%$~\cite{Won:2011ku} for the singly-Cabibbo-suppressed \DToEtapPi\ decay 
and $(-2.2\pm 2.2\pm 0.6)\%$~\cite{Onyisi:2013bjt} for the Cabibbo-favoured \DsToEtapPi\ decay, respectively. 

In this Letter the first analysis of \DcandToEtapPi\ decays at a hadron collider is presented, 
using proton-proton ($pp$) collision data corresponding to an integrated
luminosity of approximately 3\invfb, collected by the LHCb experiment. 
This allows for the large charm yields available at the LHC to be exploited, 
resulting in the most precise measurement of \CP asymmetries 
in these decays to date.

\section{Method}
\label{sec:method}

The \CP asymmetries \ACP\ are determined from the measured (raw) asymmetries 
\begin{align}
\label{e:ACPsignal}
\ACPraw(\Dcand\rightarrow f^{\pm}) 
 = \frac{N(\DcandP\rightarrow f^+) - N(\DcandM\rightarrow f^-)}{N(\DcandP\rightarrow f^+) + N(\DcandM\rightarrow f^-)} , 
\end{align}
where $N$ denotes the observed yield for the decay to a given charged final state $f^{\pm}$. 
The measured asymmetries include additional contributions other than $\ACP(\Dcand\rightarrow f^{\pm})$. 
For small asymmetries, it is possible to approximate to first order 
\begin{align}
\ACPraw \approx \ACP + \ACPprod+ \ACPdet, 
\end{align}
where \ensuremath{\ACPprod} is the asymmetry in the production of 
\ensuremath{\D^{\pm}_{(s)}} mesons in high-energy $pp$ collisions
in the \lhcb\ acceptance, and \ensuremath{\ACPdet} arises from
the difference in detection efficiencies between positively and negatively charged hadrons. 

These effects are studied using control decay modes for which \ACP\ is known precisely. 
The control decays, which have similar decay topologies as the signal decays, are the 
Cabibbo-favoured \DToKsPi\ and \DsToPhiPi\ decays for \DToEtapPi\ and \DsToEtapPi, respectively. 
The \CP\ asymmetries in these control decays have been measured at the $10^{-3}$ level  
by the \belle\ and \dzero\ collaborations~\cite{Ko:2012pe,Abazov:2013woa}.

The differences between the \CP\ asymmetries measured in the \DcandToEtapPi\ decays and 
in the corresponding control channels are defined as 
{\setlength\arraycolsep{0.1em}
\begin{eqnarray}
\label{eq:deltaformula}
\DeltaACP(\DToEtapPi) &\equiv& \ACP(\DToEtapPi) - \ACP(\DToKsPi)\\ 
                      &=& \ACPraw(\DToEtapPi) - \ACPraw(\DToKsPi) + \ACPKzKzb, \nonumber\\
\DeltaACP(\DsToEtapPi)&\equiv& \ACP(\DsToEtapPi) - \ACP(\DsToPhiPi) \nonumber\\
                      &=& \ACPraw(\DsToEtapPi) - \ACPraw(\DsToPhiPi). \nonumber
\end{eqnarray}
}These equations assume that the kinematic distributions of the pion and of the $D_{(s)}$ meson are similar 
in the signal and control channels, so that detection and production asymmetries largely cancel in the difference. 
The uncertainty associated to this assumption is discussed in Sec.~\ref{sec:Fit}.
The $\ACPKzKzb$ term in Eq.~\ref{eq:deltaformula} represents the kaon asymmetry 
in \DToKsPi\ decays, which arises from regeneration and from mixing and \CP violation in the $\Kzb-\Kz$ system.
This contribution is estimated using simulations, as described in Ref.~\cite{LHCB-PAPER-2014-013}, 
to be $(-0.08\pm 0.01)\%$. 
The \CP\ asymmetry in the singly-Cabibbo-suppressed  \DToEtapPi\ decay is therefore given by 
\begin{equation}
\ACP(\DToEtapPi) \approx \DeltaACP(\DToEtapPi) + \ACP(\DToKsPi). 
\end{equation}
Similarly, the \CP asymmetry for the Cabibbo-favoured \DsToEtapPi\ decay is approximated as 
\begin{equation}
\ACP(\DsToEtapPi) \approx \DeltaACP(\DsToEtapPi) + \ACP(\DsToPhiPi). 
\end{equation}

\section{Detector}
\label{sec:Detector}

The \lhcb detector~\cite{Alves:2008zz,LHCb-DP-2014-002} is a single-arm forward
spectrometer covering the \mbox{pseudorapidity} range $2<\eta <5$,
designed for the study of particles containing \bquark or \cquark
quarks. The detector includes a high-precision tracking system
consisting of a silicon-strip vertex detector surrounding the $pp$
interaction region, a large-area silicon-strip detector located
upstream of a dipole magnet with a bending power of about
$4{\mathrm{\,Tm}}$, and three stations of silicon-strip detectors and straw
drift tubes placed downstream of the magnet.
The polarity of the dipole magnet is reversed periodically throughout data taking.
The configuration with the magnetic field vertically upwards 
(downwards) bends positively (negatively)
charged particles in the horizontal plane towards the centre of the LHC.
The tracking system provides a measurement of momentum, \ptot, of charged particles with
a relative uncertainty that varies from 0.5\% at low momentum to 1.0\% at 200\gevc.
The minimum distance of a track to a primary vertex (PV), the impact parameter (IP),
is measured with a resolution of $(15+29/\pt)\mum$,
where \pt is the component of the momentum transverse to the beam, in\,\gevc.
Different types of charged hadrons are distinguished using information
from two ring-imaging Cherenkov detectors.
Photons, electrons and hadrons are identified by a calorimeter system consisting of
scintillating-pad and preshower detectors, an electromagnetic
calorimeter and a hadronic calorimeter. Muons are identified by a
system composed of alternating layers of iron and multiwire
proportional chambers.
The online event selection is performed by a trigger,
which consists of a hardware stage, based on information from the calorimeter and muon
systems, followed by a software stage, which applies a full event reconstruction.
At the hardware trigger stage, events are required to have a muon with high \pt or a
hadron, photon or electron with high transverse-energy deposit in the calorimeters. 

In the simulation, $pp$ collisions are generated using
\pythia~6.4~\cite{Sjostrand:2006za} with a specific \lhcb
configuration~\cite{LHCb-PROC-2010-056}.  Decays of hadronic particles
are described by \evtgen~\cite{Lange:2001uf}, in which final-state
radiation is generated using \photos~\cite{Golonka:2005pn}. The
interaction of the generated particles with the detector, and its
response, are implemented using the \geant
toolkit~\cite{Allison:2006ve, *Agostinelli:2002hh} as described in
Ref.~\cite{LHCb-PROC-2011-006}.

\section{Reconstruction and sample composition}
\label{sec:Selection}

The data correspond to an integrated luminosity of approximately 3\invfb recorded in $pp$ collisions at 
centre-of-mass energies of $\sqrt{s}=7$\tev ($1$\invfb) and 8\tev ($2$\invfb).
Approximately 50\% of the data were collected in each configuration of magnet polarity.
The \ACPraw\ measurements are performed separately for the two field polarities 
and the two centre-of-mass energies. 

The signal \DcandToEtapPi\ candidates, as well as control \DToKsPi\ and \DsToPhiPi\ candidates, 
are reconstructed through the intermediate resonance decays \EtapPPG, 
$\KS\rightarrow\pip\pim$, and $\phi\rightarrow\Kp\Km$. 
The sample is divided into three mutually exclusive subsamples according to the fulfilled 
hardware trigger requirements. 
The first subsample, {T1}, consists of events for which the trigger decision is based on the 
transverse energy deposited in the hadronic calorimeter by a charged particle from the decay 
of the $\eta'$, \KS, or $\phi$ meson. 
The second subsample, {T2}, 
consists of the subset of the remaining events for which a particle 
other than the decay products of the \Dcand\ candidate is associated with a high transverse-energy deposit 
in the hadronic calorimeter.
The third subsample, {T3}, consists of the events 
accepted because of a high transverse-energy deposit in the electromagnetic 
calorimeter or a  high-\pt muon, not associated with the \Dcand\ decay and 
not included in the other subsamples. 
The hardware trigger selections do not rely on information associated with 
the same-charge pion from the \Dcand\ decay. 

One or more of the charged decay products from the $\eta'$, \KS, or $\phi$ meson is required to activate
the first stage of the software trigger, which selects a sample with enhanced heavy-flavour content by requiring 
the presence of a large-IP charged particle with $\pt >1.6\,\gevc$ ($\pt >1.7\,\gevc$) 
in the $8\tev$ ($7\tev$) data. 
In the second stage of the software trigger, each selected event is required to contain at least one 
combination of three tracks that meet loose requirements on the IP of the final-state particles and 
on the invariant mass of the charged decay products. 

\begin{figure*}[tb]
  \begin{center}
    \includegraphics[width=0.49\linewidth]{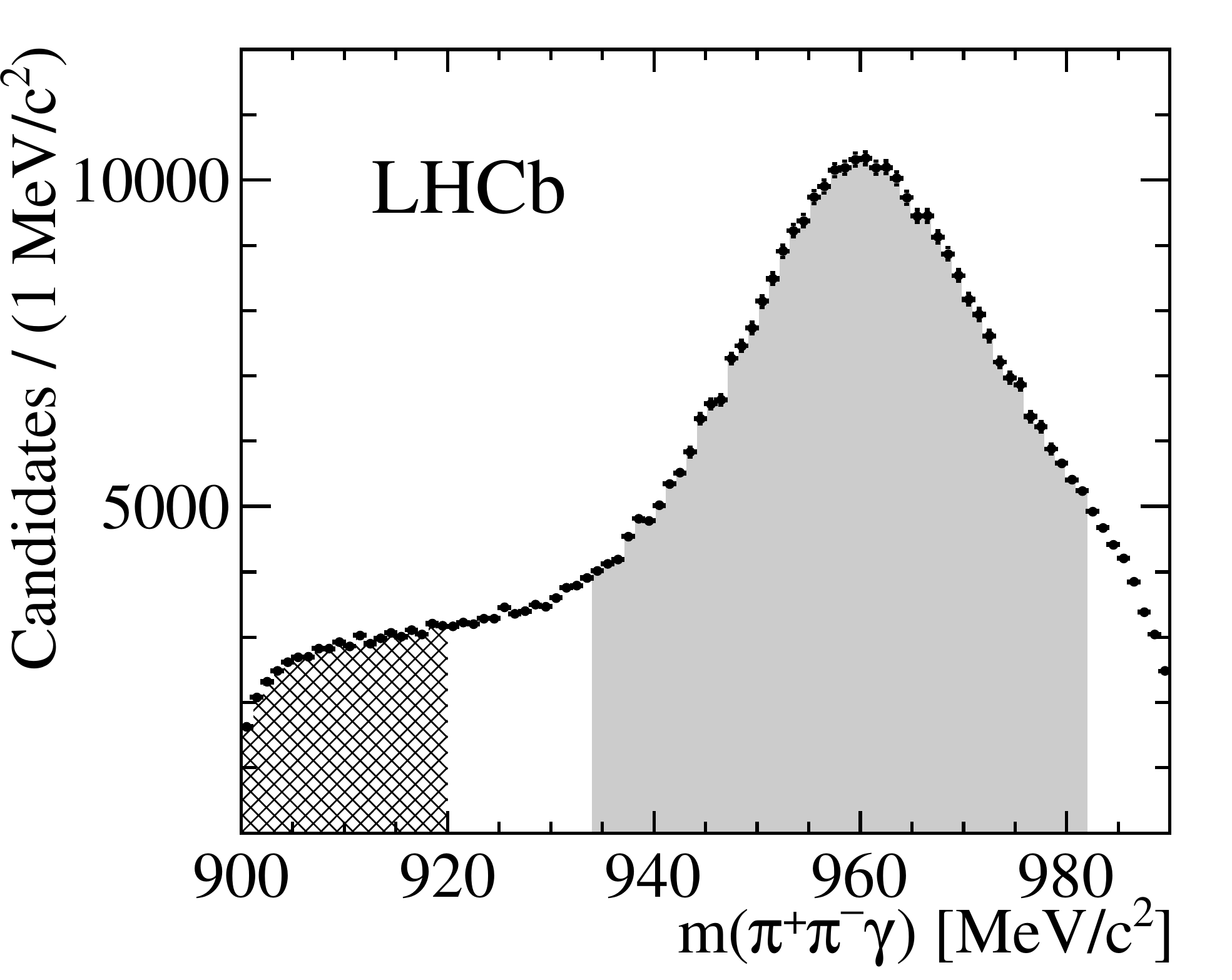}\put(-40,140){(a)} 
    \includegraphics[width=0.49\linewidth]{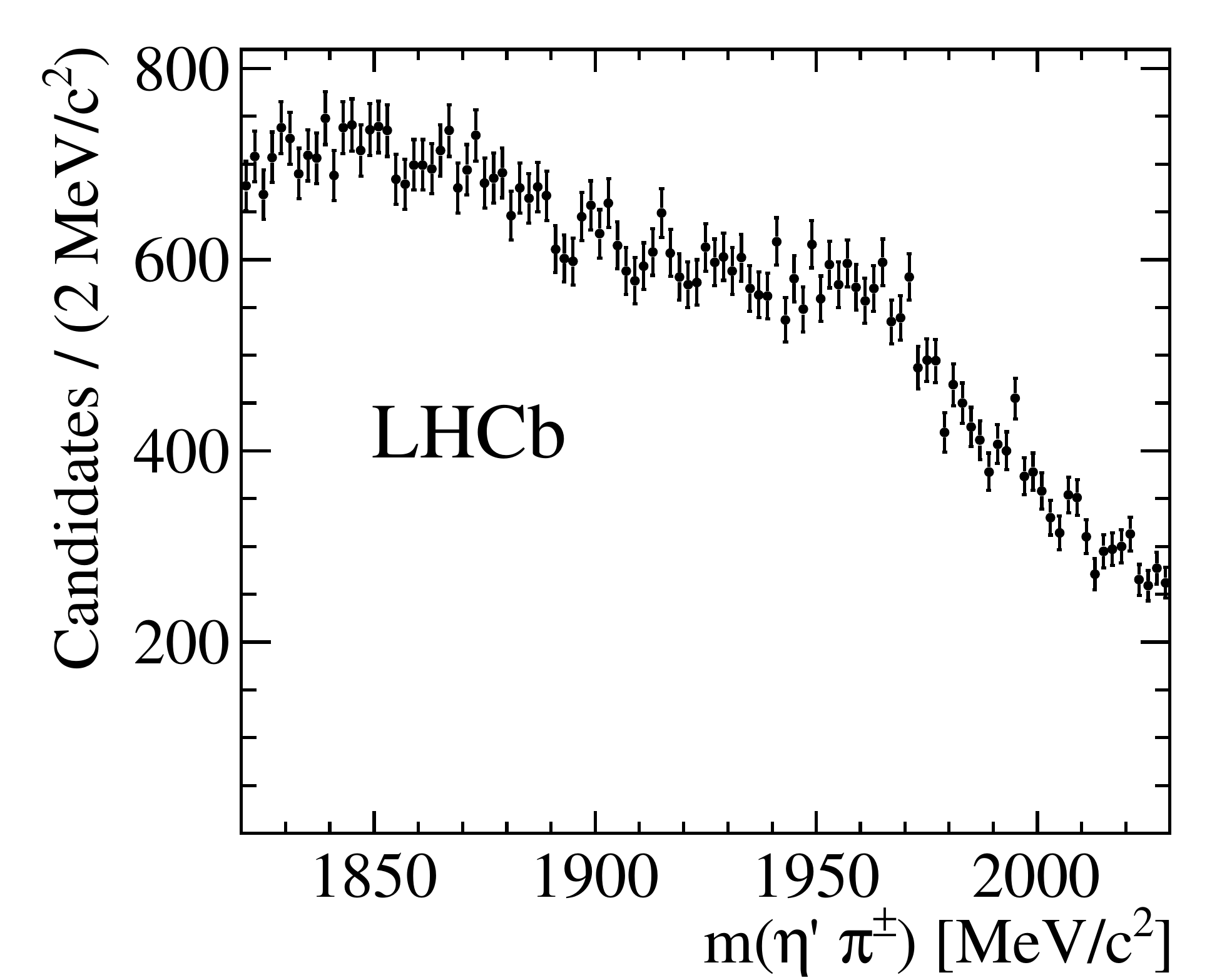}\put(-40,140){(b)} 
    \vspace*{-0.5cm}
  \end{center}
  \caption{
    Distribution of (a)~$m(\pip\pim\gamma)$ for $\Dcand \to \eta'\pipm$ candidates. 
    The grey solid 
    area represents the signal region, while the black hatched 
    area represents 
    the $m(\pip\pim\gamma)$ sideband. 
    Distribution of (b)~$m(\eta'\pipm)$ for $\Dcand \to \eta'\pipm$ candidates in the $m(\pip\pim\gamma)$ sideband. 
  \label{fig:etapMass}}
\end{figure*}

For the \DcandToEtapPi\ channels, the $\eta'$ candidates are reconstructed 
by combining pairs of oppositely charged particles with a photon of $\pt >1\,\gevc$. 
The \Etap\ charged decay products must not be identified as kaons by the particle identification system~\cite{LHCb-DP-2014-002}, 
and must be displaced from the PV. 
Photon candidates are reconstructed from clusters of energy deposits in the electromagnetic calorimeter.
The absence of tracks pointing to the energy-cluster barycentre is used 
to distinguish neutral from charged particles. 
For high-$\pt $ photons a multivariate algorithm based on the 
shape parameters of the cluster %
is used to reject $\pi^0\rightarrow\gamma \gamma$ background in which the two photons are
reconstructed as a single cluster~\cite{LHCb-DP-2014-002}. 
To maximize sensitivity to \ACPraw(\DcandToEtapPi), the three-particle mass is required to satisfy 
$0.934<m(\pip\pim\gamma)<0.982\,\gevcc$, as shown in Fig.~\ref{fig:etapMass}(a) by the light-shaded region. 

The \KS\ candidates are formed from a pair of non-prompt, oppositely charged 
high-momentum particles reconstructed in the vertex detector. 
A good-quality vertex fit 
and sufficient separation from the
PV are required for the decay vertex of the \KS\ candidate. 
The $\pip\pim$ mass is required to lie in the range 
$0.4626<m(\pip\pim)<0.5326\,\gevcc$. 

To reconstruct $\phi$ candidates,  two oppositely charged, 
large-IP particles, 
classified as kaons by the particle identification system~\cite{LHCb-DP-2014-002}, are combined. 
The $K^+K^-$ mass is required
to be within $\pm 20\,\mevcc$ of the known $\phi$ mass~\cite{PDG2014}.

Selected $\eta'$, \KS, and $\phi$ candidates are combined 
with a third non-prompt charged pion (bachelor particle) to form a \Dcand\ candidate. 
The selection criteria for the bachelor pion 
are chosen to be as similar as possible between signal and control samples. 
To suppress background contributions from $\Dcand\rightarrow X \ell^{\pm} \nu$ and $\Dcand\rightarrow X \Kpm$ decays, 
with $X = \eta'$, $\phi$, or \KS, the bachelor particle  
must be identified as a pion rather than as an electron, muon or kaon. 
The lepton veto removes more than $95\%$ of the electrons and muons and $9\%$ of the pions, and the 
kaon veto rejects about $95\%$ of the kaons while retaining $90\%$ of all pions~\cite{LHCb-DP-2014-002}.
Fiducial requirements are imposed to exclude kinematic regions 
where reconstruction and particle identification of the bachelor pion suffer 
from large charge-dependent asymmetries~\cite{Aaij:2011in}.

Candidate \Dcand\ mesons are required to have $\pt>2\,\gevc$ in all decay modes, and mass in the range
$1.82 < m(\eta^\prime \pipm)< 2.03\,\gevcc$ for the signal \DcandToEtapPi\ decays and 
$1.80 < m(\KS \pipm)< 2.03\,\gevcc$ ($1.80 < m(\phi \pipm)< 2.03\,\gevcc$) for the  
\DToKsPi\ (\DandDsToPhiPi) control mode. 
To calculate the \Dcand\ mass~\cite{Hulsbergen:2005pu}, the \Etap\ candidate mass is constrained to its known 
value~\cite{PDG2014}, without placing constraints on the origin of the \Dcand\ meson. 
The charged decay products of the reconstructed \Dcand\ 
candidates are required to match one of the three-track combinations that activated the second stage 
of the software trigger. 
The scalar sum of the transverse momenta of charged decay products must exceed $2.8\,\gevc$ for all decay modes. 
In events with multiple \Dcand\ candidates only
one randomly selected candidate is kept. This procedure removes less than $2\%$ of the original candidates. 

A combinatorial background contribution is present in all decay modes. 
Background from partially reconstructed $\Dspm\to\eta'\rho^{\pm}$ decays is suppressed by requiring 
$m(\eta^\prime \pipm)>1.82\,\gevcc$. 
Background from \DcandToPiPiPi\ decays, paired with a random photon, is suppressed by requiring the 
invariant mass of the three charged hadrons to be less than $1.80\,\gevcc$. 
A contribution from $\DandDsToPhiPi$ decays, with $\phi\rightarrow\pip\pim\pi^0$ (denoted below as $\DToPhiPiPPP$), is also 
present, where one of the photons in the $\pi^0\to \gamma\gamma$ decay is not reconstructed  
or the two photons are reconstructed as a single cluster. 

Background from $\Dspm\rightarrow \KS\Kpm$ and $\Dspm\rightarrow \KS\pipm\pi^0$ decays 
($\Dspm\rightarrow\phi\pipm\pi^0$ and non-resonant $\Dspm\rightarrow\Kp\Km\pipm$ decays), 
where the bachelor kaon is misidentified as a pion or the $\pi^0$ is not reconstructed, 
contributes negligibly to the \DToKsPi\ (\DsToPhiPi) candidate mass spectrum. 

The \DcandToEtapPi\ candidates originating from the decays of \bquark\ hadrons 
are suppressed by requiring a good quality of the \Dcand\ vertex fit, 
performed with the origin of the \Dcand\ constrained to the associated PV but 
without a constraint on the \Etap\ candidate mass.  
Non-prompt \DsToPhiPi\ and \DToKsPi\ candidates 
are rejected by requiring a small difference between the quality 
of the fit of the PV formed with and without 
the tracks assigned to the reconstructed \Dcand\ candidate. 

\section{Determination of the asymmetries}
\label{sec:Fit}

For each final state, the data are divided into twelve mutually exclusive subsamples, 
according to the two $pp$ centre-of-mass energies, two magnet polarities,
and three hardware trigger selections. 
Since detection asymmetries depend on the kinematic properties of the process under study, 
\Dcand\ candidates in each subsample are divided into $3 \times 3$ bins of 
transverse momentum and pseudorapidity of the bachelor pion.
The bin edges in \pt\ are defined as $0.5$, $1.5$, $3.0$, and $20.0\,\gevc$,
and the bin edges in $\eta$ are defined as $2.0$, $2.8$, $3.2$, and $5.0$. 
While the kinematic distributions of the bachelor pion for the signal and \DsToPhiPi\ control decays are in good agreement, 
the average bachelor-pion \pt\ ($\eta$) is $30\%$ lower ($5\%$ higher) in the \DToKsPi\ control channel. 
The binning reduces the effect of the discrepancies between the bachelor-pion kinematic distributions 
for signal and control decays, thus improving the suppression of $\ACPdet$ in the differences of raw asymmetries.
For each of the twelve subsamples, the raw \CP\ asymmetries of the \DcandToEtapPi\ signal channels are determined with 
a maximum likelihood fit to the unbinned $\eta'\pi$ invariant mass distribution,
performed simultaneously for positively and negatively charged \Dcand\ candidates, and for the 
nine $\pt-\eta$ bins. 

\begin{figure*}[tb]
  \begin{center}
    \includegraphics[width=0.49\linewidth]{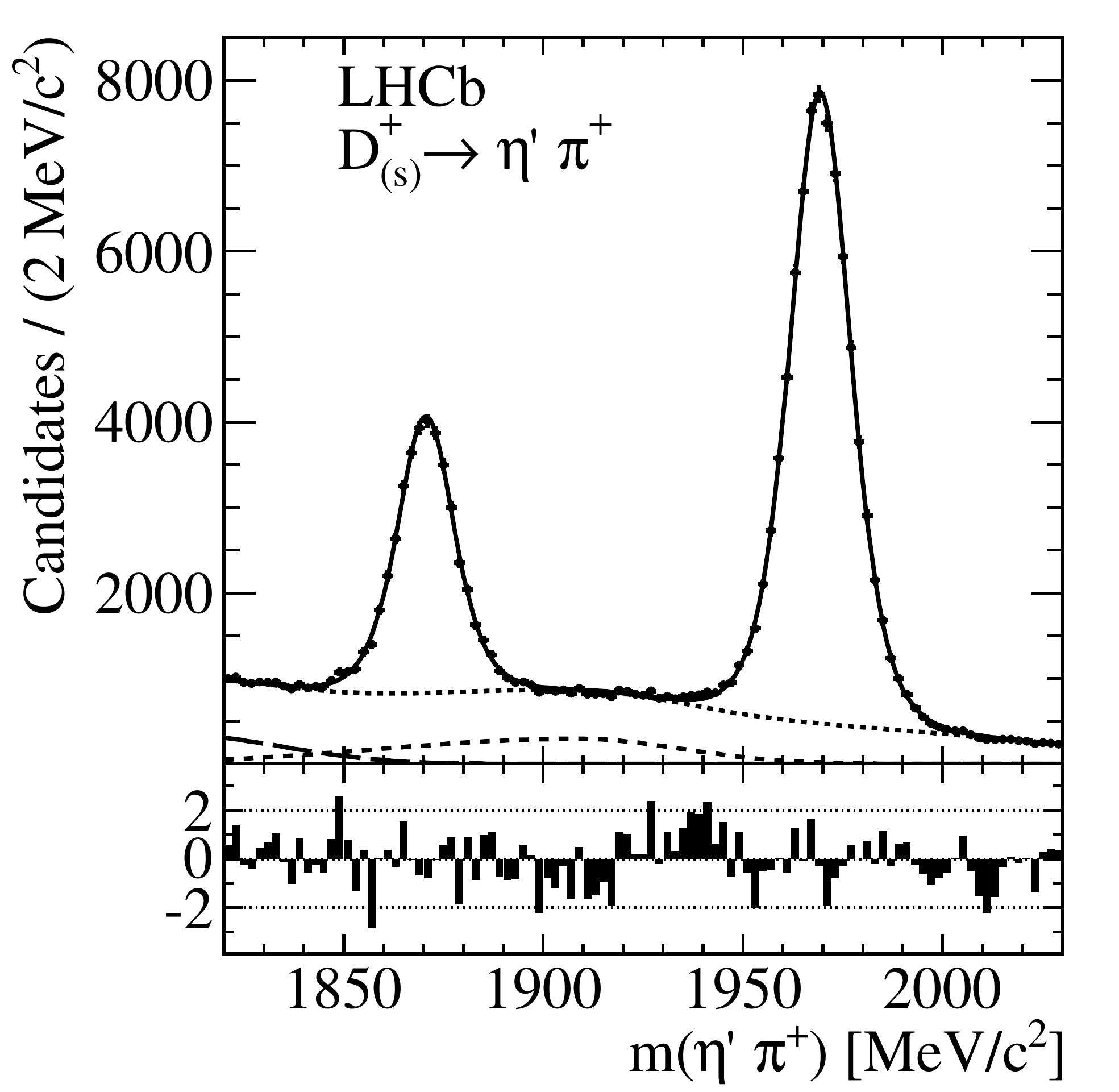}\put(-40,190){(a)} 
    \includegraphics[width=0.49\linewidth]{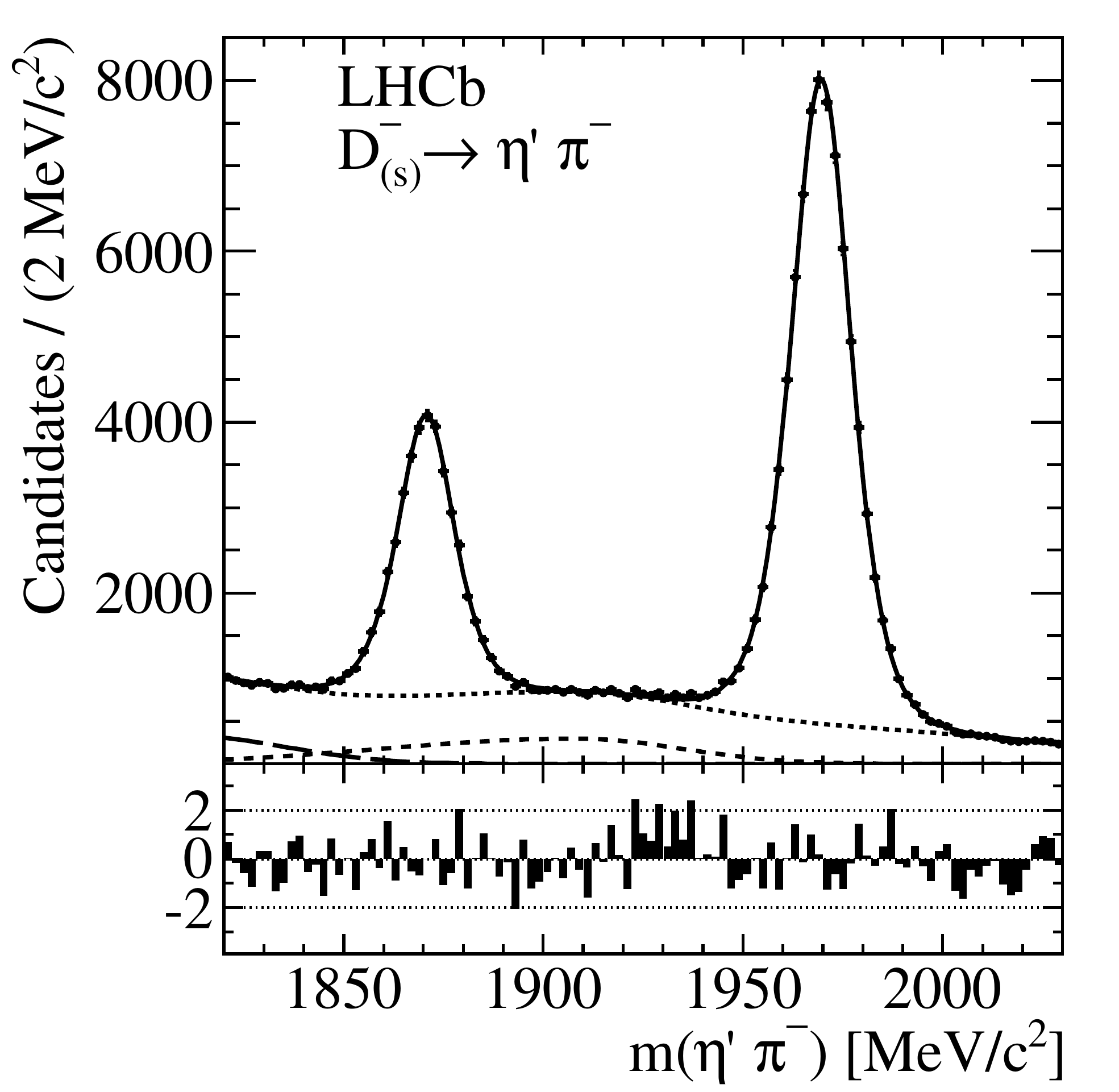}\put(-40,190){(b)} 
    \vspace*{-0.5cm}
  \end{center}
  \caption{
  Mass distribution of $\EtapPi^\pm$ candidates, combined over all kinematic bins, $pp$ centre-of-mass energies, 
  and hardware trigger selections, for (a)~positively and (b)~negatively charged \Dcand\ candidates. 
  Points with errors represent data, while the curves represent the fitted model (solid), 
  the \DspmToPhiPiPPP\ (dashed) and \DpmToPhiPiPPP\ (long-dashed) components, and the sum of all background 
  contributions (dotted), including combinatorial background.  
  Residuals divided by the corresponding uncertainty are shown under each plot.
  }
  \label{fig:dmass}
\end{figure*}

The fit model comprises two signal components for the \Dpm and \Dspm\ resonances, a combinatorial background component, 
and two peaking components accounting for background from \DToPhiPiPPP\ decays.
The signal components are modelled by Johnson $S_U$ distributions~\cite{Johnson:1949zj}: 
\begin{equation}
 f(x; \mu,\sigma,\delta,\gamma) \propto  \left[1+\left(\frac{x-\mu}{\sigma}\right)^2\right]^{-\frac{1}{2}} \exp\left\{-\frac{1}{2}\left[\gamma + \delta\sinh^{-1}\left(\frac{x-\mu}{\sigma}\right)\right]^2\right\}. 
\end{equation}
The parameters $\mu$ and $\sigma$, which govern the mean and width of each distribution, 
are fitted independently for \Dpm\ and \Dspm, 
and can vary with the charge and pseudorapidity of the bachelor pion. 
The remaining two parameters, $\delta$ and $\gamma$, characterising the tails of the Johnson $S_U$ distributions, are 
common between the two signal components and 
are required to be the same across all $\pt -\eta$ bins, but can vary with the charge of the bachelor pion. 
The combinatorial background is parametrised by a fourth-order polynomial, %.
whose parameters can vary independently for positive and negative charges 
and for different bins in the bachelor-pion pseudorapidity. 
The parameters of the background model, for each charge and each bin in pseudorapidity of the bachelor pion,
are Gaussian-constrained to the results of fits of the same functional form to the corresponding 
$m(\eta^\prime\pipm)$ distributions from the $m(\pip\pim\gamma)$ sideband (Fig.~\ref{fig:etapMass}(b)). 
The \DToPhiPiPPP\ background components are described by empirical functions~\cite{delAmoSanchez:2010ae} 
derived from simulated events. 
The yields and charge asymmetries of signal and combinatorial components in each $\pt -\eta$ bin, 
and the total yields of the \DToPhiPiPPP\ contributions are free parameters in the fit. 
For the \DToPhiPiPPP\  components, the raw \CP\ asymmetries and the fraction of the total yields 
in each $\pt -\eta$ bin are determined from $\Dcand\to\phi\pipm$ control decays, with $\phi\to K^+K^-$. 
The model well reproduces the charge-integrated $m(\eta^\prime\pipm)$ distributions in all $\pt$ bins. 
To estimate the goodness of fit, in each of the twelve subsamples 
the $\chi^2$ of the fitted model is calculated for the binned $m(\eta^\prime\pipm)$ distribution in all $\pt -\eta$ bins. 
The $p$-value is greater than $5\%$ in all cases.
The results of the fit to the \EtapPi\ mass distribution for \DcandToEtapPi\ candidates are 
shown in Fig.~\ref{fig:dmass}. 
The signal yields, 
combined over all kinematic bins, $pp$ centre-of-mass energies, and hardware trigger selections, 
are $N(\DToEtapPi) = (62.7\pm 0.4)\times 10^3$ and $N(\DsToEtapPi) = (152.2\pm 0.5)\times 10^3$, respectively. 

\begin{figure*}[tb]
  \begin{center}
    \includegraphics[width=0.49\linewidth]{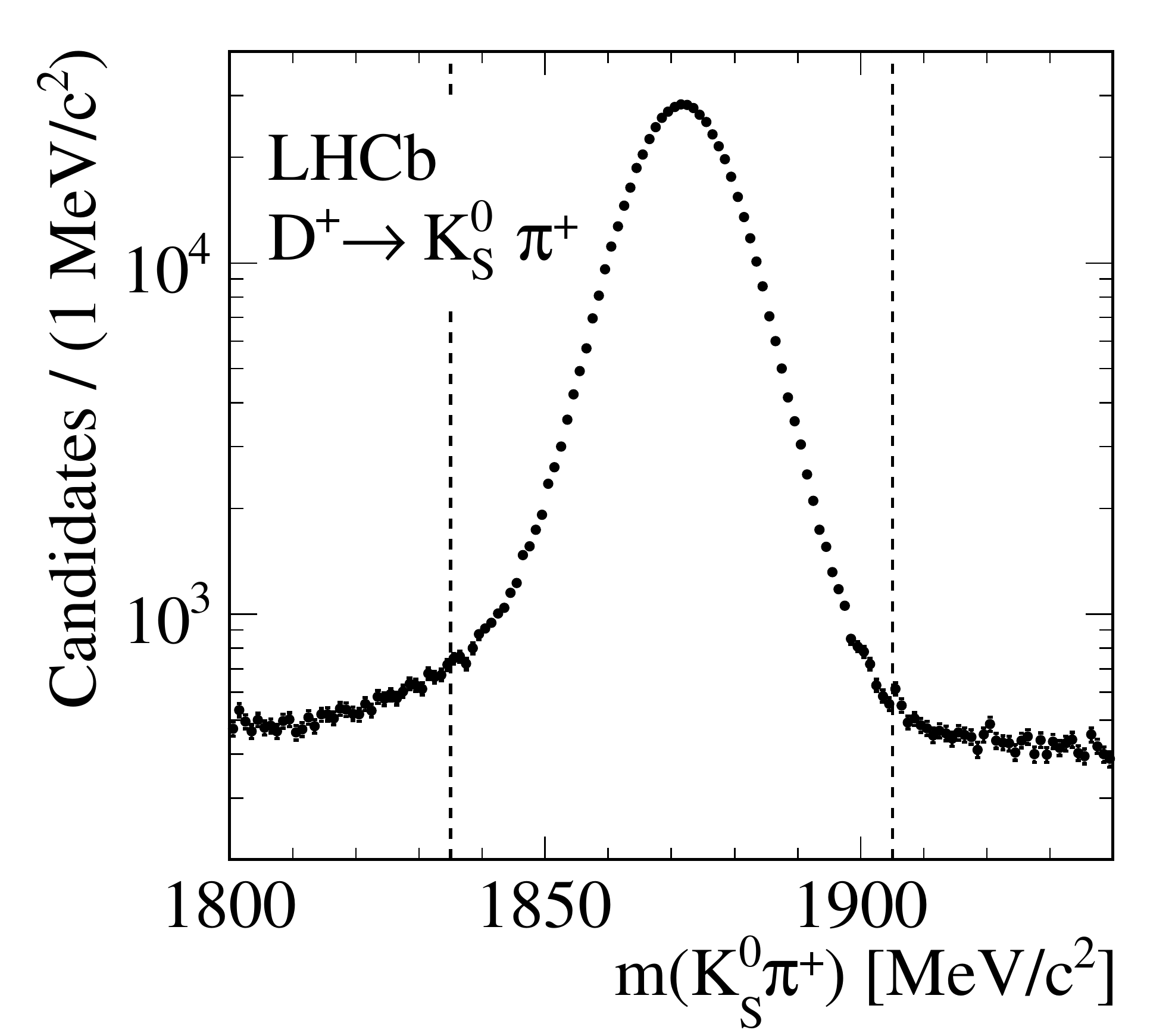}\put(-40,150){(a)} 
    \includegraphics[width=0.49\linewidth]{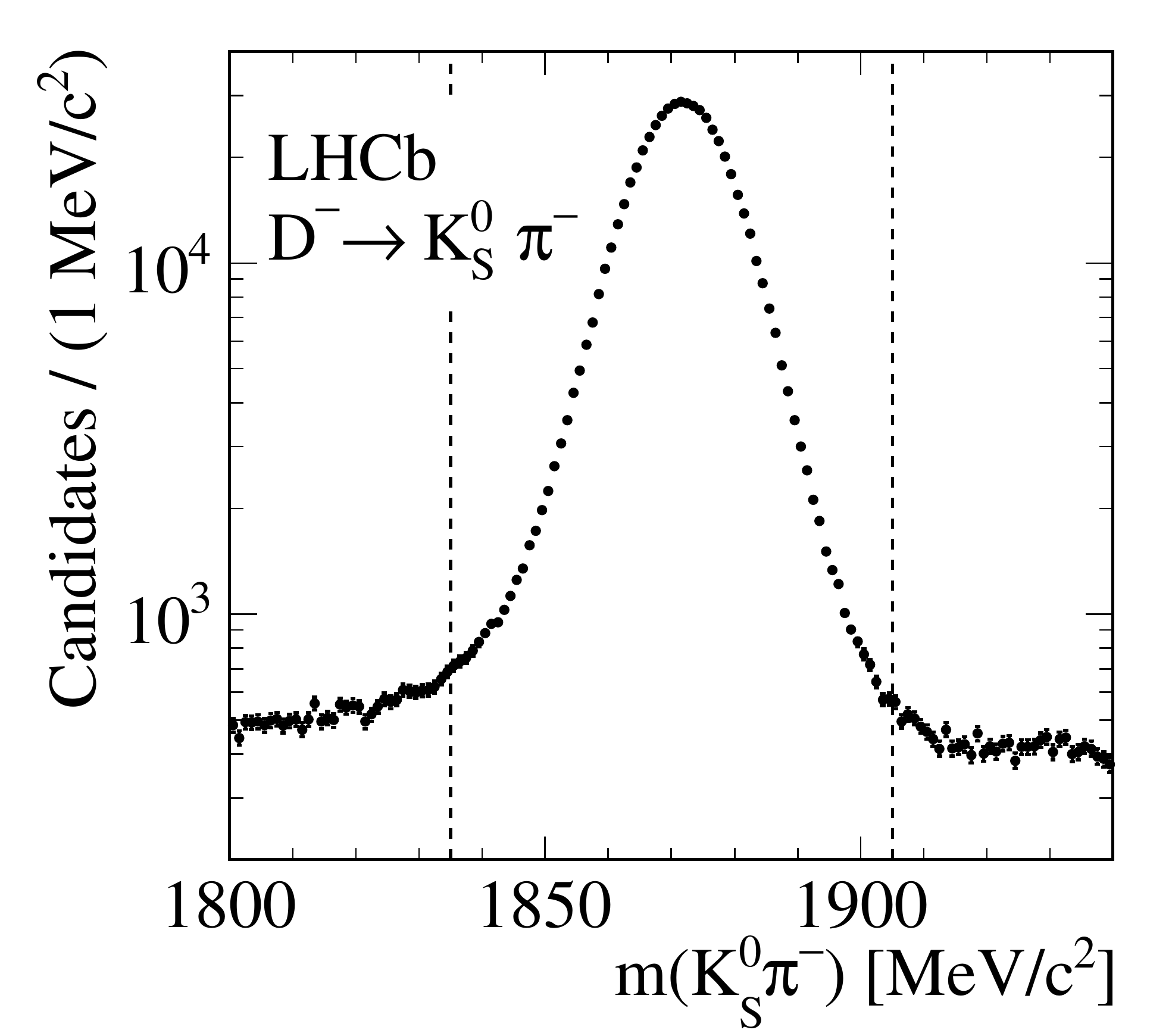}\put(-40,150){(b)} \\ 
    \includegraphics[width=0.49\linewidth]{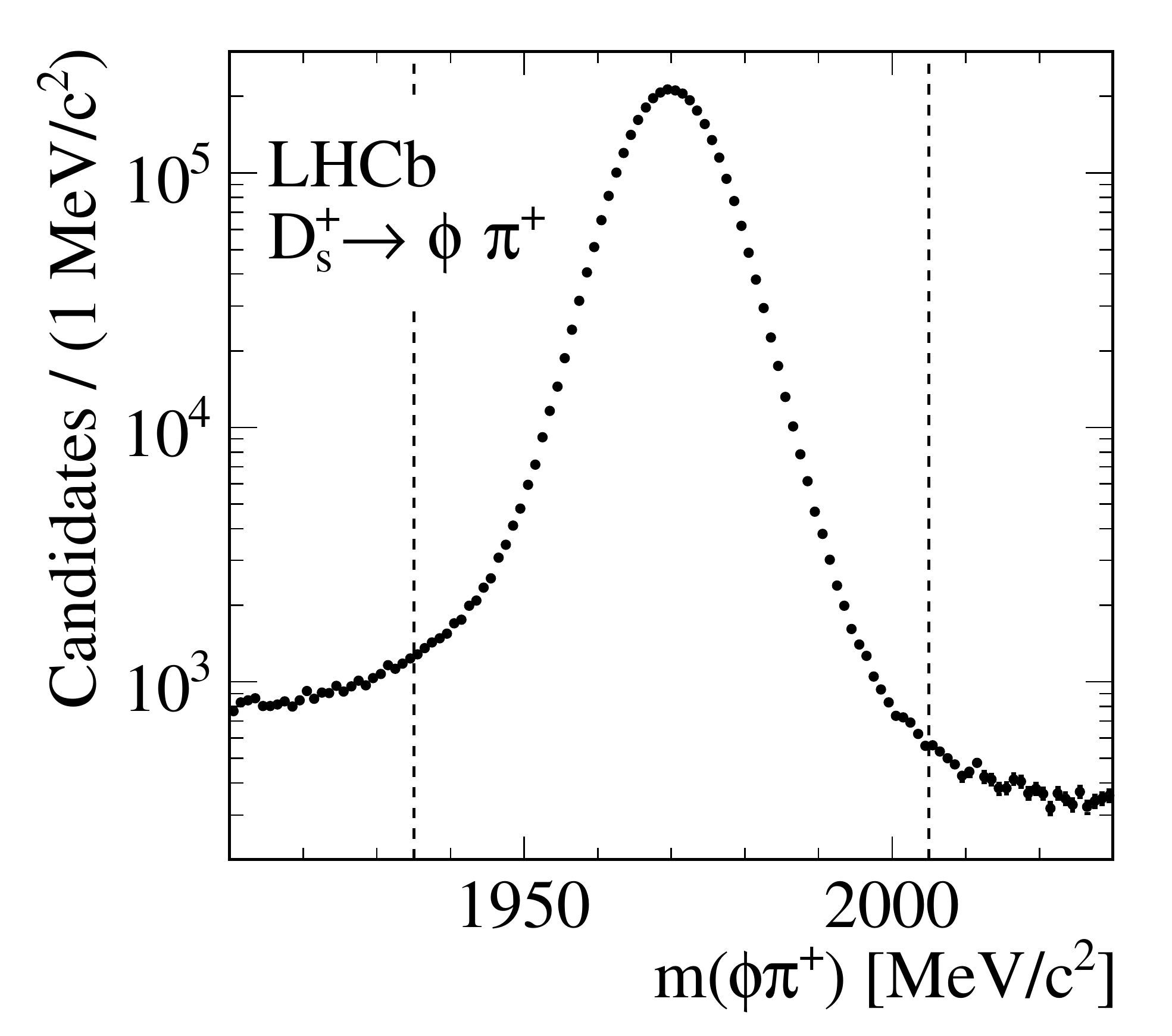}\put(-40,150){(c)} 
    \includegraphics[width=0.49\linewidth]{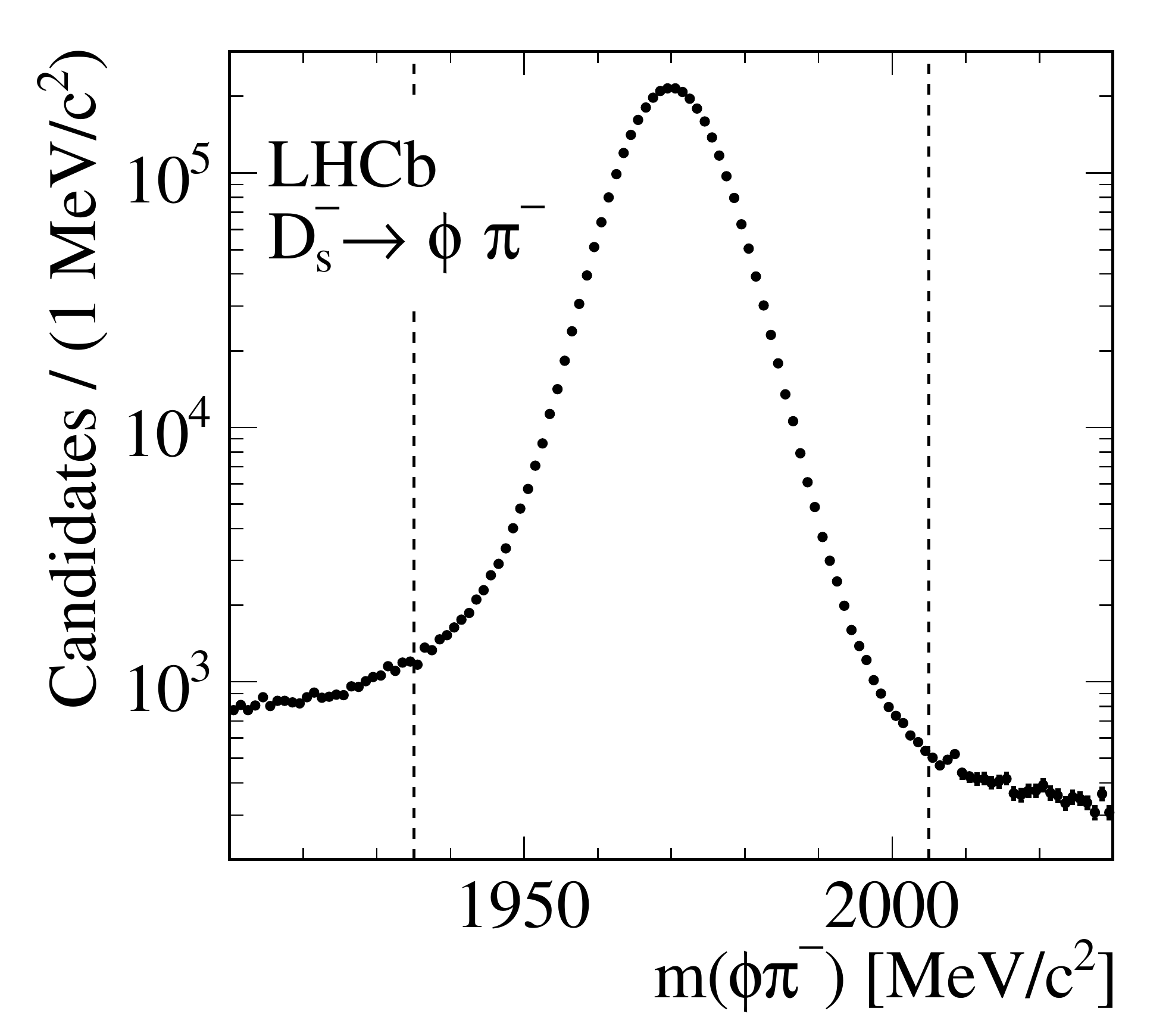}\put(-40,150){(d)} 
    \vspace*{-0.5cm}
  \end{center}
  \caption{
  Top: $\KS \pipm$ mass distribution for (a)~positively and (b)~negatively charged \Dpm\ candidates.
  Bottom: $\phi \pipm$ mass distribution for (c)~positively and (d)~negatively charged \Dspm\ candidates. 
  The signal regions are enclosed within the vertical dashed lines.
  The mass distributions are combined over all kinematic bins, $pp$ centre-of-mass energies, and hardware trigger selections. 
  }
  \label{fig:csdmass}
\end{figure*}

Due to the high purity of the control samples, 
the raw \CP\ asymmetries for the \DsToPhiPi\ and \DToKsPi\ decay modes are 
extracted by counting the numbers of positively and negatively charged candidates 
in the signal mass range and subtracting the corresponding numbers in the sidebands, shown in Fig.~\ref{fig:csdmass}. 
For the \DToKsPi\ decay, the sidebands are defined as 1.800\,--\,1.835\gevcc and 
1.905\,--\,1.940\gevcc, and the signal range as 1.835\,--\,1.905\gevcc. 
For the \DsToPhiPi\ channel the sidebands are defined as 1.910\,--\,1.935\gevcc\  
and 2.005\,--\,2.030\gevcc , and the signal range as 1.935\,--\,2.005\gevcc.  
The event yields determined in the \DsToPhiPi\ sidebands are 
scaled by a factor $1.4$ 
to account for the different widths of the sideband and signal ranges.
Background from $\Dspm\rightarrow \KS\Kpm$, $\Dspm\rightarrow \KS\pipm\pi^0$,  
$\Dspm\rightarrow\phi\pipm\pi^0$, and non-resonant $\Dspm\rightarrow\Kp\Km\pipm$ decays is neglected. 
The effect of the small fraction of \Dcand\ signal leaking into the sidebands, which may depend  
on the charge, $p_T$ and pseudorapidity of the bachelor pion, is 
considered as a source of systematic uncertainty. 

\begin{figure}[tb]
  \begin{center}
    \includegraphics[width=0.68\linewidth]{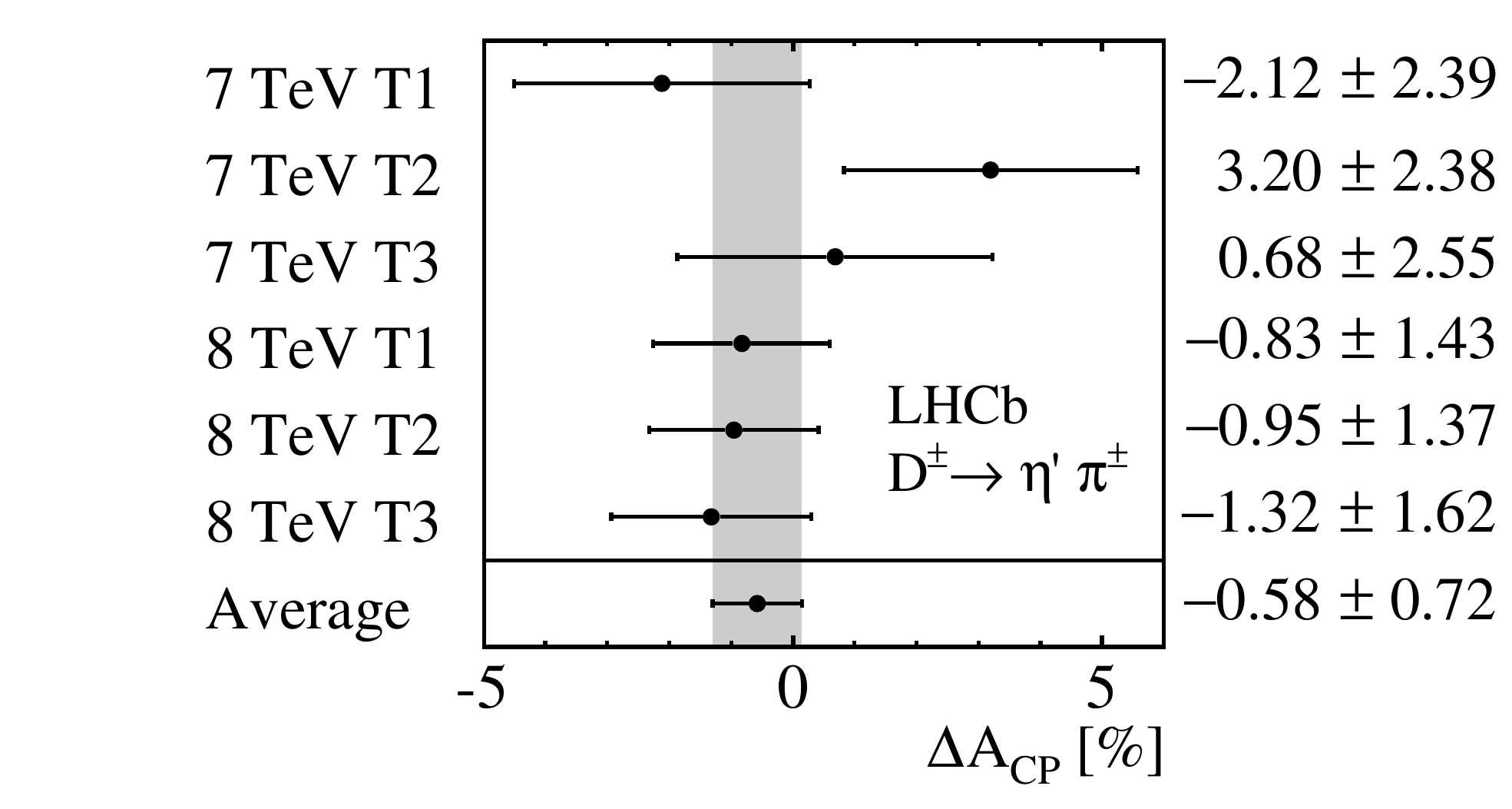}\put(-90,100){(a)}\\
    \includegraphics[width=0.68\linewidth]{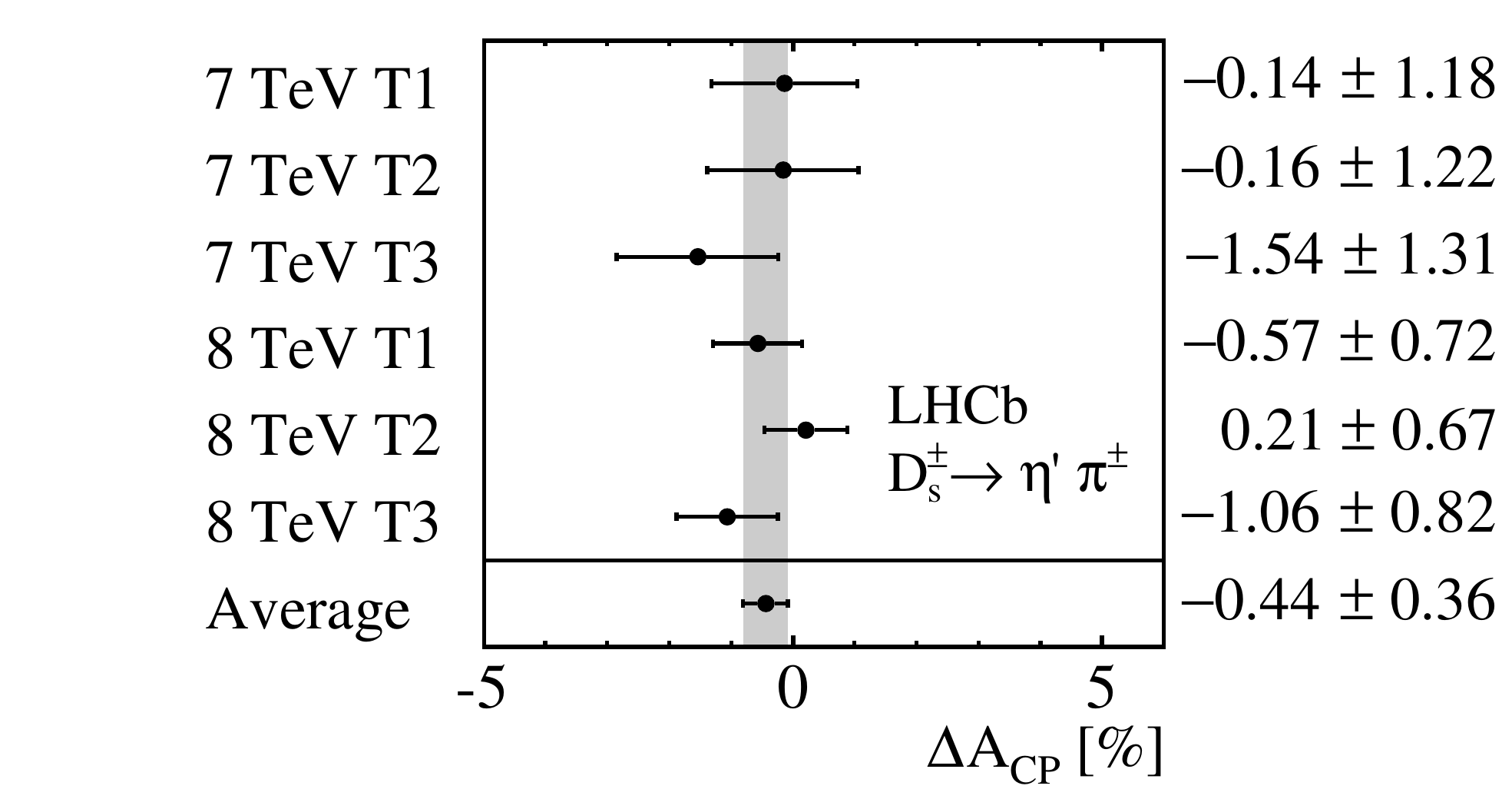}\put(-90,100){(b)}\\
    \vspace*{-0.5cm}
  \end{center}
  \caption{%\small 
  $\Delta\ACP$ results for (a)~\DToEtapPi\ and (b)~\DsToEtapPi\ decays, as a function of 
  $pp$ centre-of-mass energy and trigger selection. Uncertainties are statistical only. A shaded 
  band representing the $68.3\%$ confidence intervals obtained from the weighted average  
  over all the samples is shown to guide the eye. }
  \label{fig:summaries}
\end{figure}

For each subsample, the differences of raw asymmetries for signal and associated control channels 
are calculated in each $\pt-\eta$ bin. 
The weighted averages of the results obtained in the nine bins are then 
evaluated, taking into account the covariance matrix $V$, 
calculated as the sum of the covariance matrices for the results of 
the \DcandToEtapPi\ fit and of the sideband subtraction for control decays.     
The weights are 
$w_i = \sum_{k}V_{ik}^{-1}/\left(\sum_{j}\sum_{k}V_{jk}^{-1}\right)$, where $i$, $j$, and $k$ 
run over the $\pt-\eta$ bins. 
The resulting \DeltaACP\ values are averaged with equal weights
over the two magnet polarities. 
Detection asymmetries 
that differ between the signal and control decays are suppressed  
in this average. The results for the signal channels are shown in Fig.~\ref{fig:summaries}. 
Finally, the inverse-variance weighted average of the \DeltaACP\ values obtained 
for the two $pp$ centre-of-mass energies and the three hardware trigger selections is calculated.  
No significant charge asymmetry is observed for the combinatorial background component in any of the subsamples. 
The inverse-variance weighted average of \ACPraw\ for the combinatorial background is $(0.92\pm 0.72)\%$, where the 
error is statistical only.

\section{Systematic uncertainties}
\label{sec:Systematics}

The contributions to the systematic uncertainty on the inverse-variance weighted \DeltaACP\ average 
are described below and summarised in Table~\ref{tab:syst}. 
The overall systematic uncertainties are obtained by adding 
the individual contributions in quadrature. 

%%%%%%%%%%%%%%%%%%%%%%%%%%%%%%%
\begin{table}[b]
    \caption{Systematic uncertainties (absolute values in \%) on \DeltaACP. 
    The total systematic uncertainty is the sum in quadrature of the individual contributions.
    \label{tab:syst}}
  \begin{center}
    \begin{tabular}{l|cc} \hline
Source               & $\delta[\DeltaACP(\Dpm)]$ & $\delta[\DeltaACP(\Dspm)]$ \\ \hline
Non-prompt charm     & $0.03$   & $0.03$            \\
Trigger              & $0.09$   & $0.09$         \\
Background model     & $0.50$   & $0.19$         \\
Fit procedure        & $0.08$   & $0.04$        \\
Sideband subtraction & $0.03$ & $0.02$ \\
$\Kz$ asymmetry      & $0.08$   & $-$         \\ 
$\pi^{\pm}$ detection asymmetry & $0.06$   &  $0.01$        \\ 
$\Dcand$ production asymmetry & $0.07$       &  $0.02$      \\ \hline
Total          & $0.53$   & $0.22$         \\ \hline 
\end{tabular}
  \end{center}
\end{table}
%%%%%%%%%%%%%%%%%%%%

The selection of signal and control sample candidates removes the majority of background 
from non-prompt \Dcand\ mesons, originating from the decay of a \bquark~hadron. 
The remaining secondary \Dcand\ mesons may introduce a bias in the measured \CP\ asymmetries due 
to a difference in the production asymmetries for \bquark~hadrons and \Dcand\ mesons. 
This bias might not cancel in the difference of measured asymmetries for signal and control channels, 
due to differences in the final-state reconstruction. 
In order to investigate this bias, the \Dcand\ production asymmetries in \DcandToEtapPi\ decays 
are modified using 
$\ACPprod' = (\ACPprod+f \ACPprod^{b})/(1+f)$,
where $f$ is the fraction of secondary \Dcand\ candidates in a particular decay channel
and $\ACPprod^{b}$ is the corresponding \bquark-hadron production asymmetry.
The fraction $f$ is estimated from the measured cross-sections for inclusive production of 
$\Dpm$, $\D_s^{\pm}$, and \bquark~hadrons~\cite{LHCb-PAPER-2012-041,LHCb-PAPER-2010-002},  
the inclusive branching fractions
$\BR(b\to D^{\pm}X)$ and $\BR(b\to D^{\pm}_s X)$~\cite{PDG2014},
and the efficiencies calculated from simulation. The resulting values of $f$ are below $6\%$.
The \bquark-hadron production asymmetry $\ACPprod^{b}$ is taken from existing  
measurements for \Bpm, $B_{(s)}^{0}$, and 
\Lb hadrons~\cite{LHCB-PAPER-2011-024,LHCB-PAPER-2014-032,LHCb-PAPER-2014-042, LHCb-PAPER-2014-053,LHCb-PAPER-2014-020}.
Under the assumption that the bias due to $\ACPprod^{b}$ does not cancel in the difference 
of measured asymmetries for signal and control channels, 
the systematic uncertainty on \DeltaACP\ is evaluated by recalculating the \CP\ asymmetries 
using $\ACPprod'$ for the signal decay modes and $\ACPprod$ for the control samples. 

Potential trigger biases are studied using \DandDsToPhiPi\ decays, with $\phi\to K^+K^-$. 
The \CP\ asymmetries measured in the subsamples defined by the T2 and T3 trigger selections 
are compared to the asymmetries from the T1 subsample, which is based on charge-symmetric combinations of tracks.
No statistically significant discrepancy is observed, and the 
statistical uncertainty of the difference 
is assigned as a systematic uncertainty.
This systematic uncertainty accounts for residual trigger-induced biases in the 
difference of measured asymmetries for signal and control channels.

Different background parametrizations can change the ratio of signal and background 
and affect the observed asymmetry. 
The nominal model is modified by replacing, for all subsamples, 
the fourth-order polynomial with other empirically chosen functions, 
a second-order polynomial or an ARGUS function~\cite{Albrecht:1990am}. 
Different fit configurations are tested, in which the background parameters are fixed according to the results of a fit 
to the $m(\pip\pim\gamma)$ sideband, or in which the $\DToPhiPiPPP$ background fractions are varied.
The maximum deviations from the results of the nominal fit, observed with any of the alternative models 
providing a reasonable fit to the data, 
are assigned as systematic uncertainties. This represents the largest contribution to the systematic uncertainties 
in both channels. This estimate is in agreement with an independent assessment, based on the increased statistical 
uncertainties on \ACPraw\ when the constraints on the background parameters are removed from the nominal model.

The fitting procedure is validated with several pseudoexperiments using events simulated according 
to the fit model, varying the \ACPraw\ value used in generation.  
The sum in quadrature of the bias and of its statistical uncertainty is taken as a systematic uncertainty.

A systematic uncertainty is introduced for the background contributions neglected in the 
measurement of the raw asymmetries for the \DToKsPi\ and \DsToPhiPi\ control decays, and 
for the neglected fraction of \Dcand\ signal leaking into the sidebands. 
The effect of non-resonant $\Dspm\rightarrow\Kp\Km\pipm$ contributions to the \DsToPhiPi\ control sample 
is evaluated by observing the variation of $\DeltaACP(\DsToEtapPi)$ when the $K^+K^-$ mass is required
to be within $\pm 10\,\mevcc$ (instead of $\pm 20\,\mevcc$) of the known $\phi$ mass. 
The systematic uncertainty due to $\Dspm\rightarrow \KS\Kpm$, $\Dspm\rightarrow \KS\pipm\pi^0$, 
and $\Dspm\rightarrow\phi\pipm\pi^0$ is calculated from the estimated fraction of 
background events, assuming a negligible \CP\ violation 
and using the production asymmetries in \lhcb\ acceptance as an input. 
The difference of raw asymmetries in $\DeltaACP(\DToEtapPi)$ is corrected for the $\Kz$ asymmetry~\cite{LHCB-PAPER-2014-013}
and an associated systematic uncertainty equal to the applied correction is included.

The potential discrepancy in the bachelor pion kinematic distribution within each $\pt-\eta$ bin between signal 
and control samples, associated to the finite number of bins, might result in an incomplete 
cancellation of detection asymmetries. The discrepancy in \DeltaACP\ with respect to the nominal binning, 
resulting from using no kinematic binning, is assigned as a systematic uncertainty. 

The \Dcand\ production asymmetry may show a dependence on $\pt $ and $\eta$
of the charm meson. Therefore, the cancellation of production effects in \DeltaACP\ may 
be partial, since \Dcand\ kinematic distributions are different for signal and control channels. 
To estimate this effect, in each bin of the bachelor-pion kinematic distribution, the 
\DToKsPi\ and \DsToPhiPi\ candidates are given a 
weight depending on either the $\pt$ or the $\eta$ value of the \Dcand\ meson, to reproduce the \Dcand\ kinematic distribution 
of signal candidates. The effect on \DeltaACP\ is assigned as a systematic uncertainty.

The \DeltaACP\ results are stable when the requirements on the bachelor-pion 
particle identification and track quality are tightened, and 
when the constraints on the parameters of the combinatorial background component
are removed from  the fit to \DcandToEtapPi\ candidates.  
The stability of \DeltaACP\ is also investigated as a function of beam energy and hardware trigger decision. 
No significant dependence is observed, as shown in Fig.~\ref{fig:summaries}.

\section{Results and summary}
\label{sec:Results}

Using $pp$ collision data collected by the \lhcb\ experiment at centre-of-mass energies of $7$ and $8\,\tev$, 
the differences in \CP asymmetries between \DToEtapPi\ and \DToKsPi\ decays, and 
between \DsToEtapPi\ and \DsToPhiPi\ decays, are measured to be 
{\setlength\arraycolsep{0.1em}
\begin{eqnarray*}
\DeltaACP(\DToEtapPi) &=& (-0.58 \pm 0.72 \pm 0.53 )\%, \\
\DeltaACP(\DsToEtapPi) &=& (-0.44 \pm 0.36 \pm 0.22)\%. 
\end{eqnarray*}
}
In all cases, the first uncertainties are statistical and the second are systematic.

Using the previously measured values of the \CP\ asymmetries in control decays, 
$ \ACP(\DToKsPi) = (-0.024 \pm 0.094 \pm 0.067)\%$~\cite{Ko:2012pe} and 
$ \ACP(\DsToPhiPi) = (-0.38 \pm 0.26 \pm 0.08)\%$~\cite{Abazov:2013woa},
the individual \CP\ asymmetries are found to be
{\setlength\arraycolsep{0.1em}
\begin{eqnarray*}
\ACP(\DToEtapPi) &=& (-0.61\pm 0.72 \pm 0.53 \pm 0.12)\%,  \\
\ACP(\DsToEtapPi) &=& (-0.82\pm 0.36 \pm 0.22 \pm 0.27)\%,
\end{eqnarray*}
}
where the last contribution to the uncertainty comes from the 
$\ACP(\DToKsPi)$ and $\ACP(\DsToPhiPi)$ measurements. 

The measured values show no evidence of \CP\ violation, and are consistent with 
SM expectations~\cite{Cheng:2010ry,Cheng:2012wr,Li:2012cfa} 
and with previous results obtained in $e^+e^-$ collisions~\cite{Onyisi:2013bjt,Won:2011ku}. 
The results represent the most precise measurements of these quantities to date.

\section*{Acknowledgements}

\noindent We express our gratitude to our colleagues in the CERN
accelerator departments for the excellent performance of the LHC. We
thank the technical and administrative staff at the LHCb
institutes. We acknowledge support from CERN and from the national
agencies: CAPES, CNPq, FAPERJ and FINEP (Brazil); NSFC (China);
CNRS/IN2P3 (France); BMBF, DFG and MPG (Germany); INFN (Italy); 
FOM and NWO (The Netherlands); MNiSW and NCN (Poland); MEN/IFA (Romania); 
MinES and FASO (Russia); MinECo (Spain); SNSF and SER (Switzerland); 
NASU (Ukraine); STFC (United Kingdom); NSF (USA).
We acknowledge the computing resources that are provided by CERN, IN2P3 (France), KIT and DESY (Germany), INFN (Italy), SURF (The Netherlands), PIC (Spain), GridPP (United Kingdom), RRCKI and Yandex LLC (Russia), CSCS (Switzerland), IFIN-HH (Romania), CBPF (Brazil), PL-GRID (Poland) and OSC (USA). We are indebted to the communities behind the multiple open 
source software packages on which we depend.
Individual groups or members have received support from AvH Foundation (Germany),
EPLANET, Marie Sk\l{}odowska-Curie Actions and ERC (European Union), 
Conseil G\'{e}n\'{e}ral de Haute-Savoie, Labex ENIGMASS and OCEVU, 
R\'{e}gion Auvergne (France), RFBR and Yandex LLC (Russia), GVA, XuntaGal and GENCAT (Spain), Herchel Smith Fund, The Royal Society, Royal Commission for the Exhibition of 1851 and the Leverhulme Trust (United Kingdom).

\addcontentsline{toc}{section}{References}
\setboolean{inbibliography}{true}
\bibliographystyle{LHCb}
\bibliography{main,DToEtapPi,LHCb-PAPER,LHCb-CONF,LHCb-DP,LHCb-TDR}

\clearpage

\centerline{\large\bf LHCb collaboration}
\begin{flushleft}
\small
R.~Aaij$^{40}$,
B.~Adeva$^{39}$,
M.~Adinolfi$^{48}$,
Z.~Ajaltouni$^{5}$,
S.~Akar$^{6}$,
J.~Albrecht$^{10}$,
F.~Alessio$^{40}$,
M.~Alexander$^{53}$,
S.~Ali$^{43}$,
G.~Alkhazov$^{31}$
% ...

{\footnotesize \it
% ...
$ ^{5}$Clermont Universit{\'e}, Universit{\'e} Blaise Pascal, CNRS/IN2P3, LPC, Clermont-Ferrand, France\\
$ ^{6}$CPPM, Aix-Marseille Universit{\'e}, CNRS/IN2P3, Marseille, France\\
% ...
$ ^{10}$Fakult{\"a}t Physik, Technische Universit{\"a}t Dortmund, Dortmund, Germany\\
% ...
$ ^{31}$Petersburg Nuclear Physics Institute (PNPI), Gatchina, Russia\\
% ...
$ ^{39}$Universidad de Santiago de Compostela, Santiago de Compostela, Spain\\
$ ^{40}$European Organization for Nuclear Research (CERN), Geneva, Switzerland\\
% ...
$ ^{43}$Nikhef National Institute for Subatomic Physics, Amsterdam, The Netherlands\\
% ...
$ ^{48}$H.H. Wills Physics Laboratory, University of Bristol, Bristol, United Kingdom\\
% ...
$ ^{53}$School of Physics and Astronomy, University of Glasgow, Glasgow, United Kingdom\\
% ...
}
\end{flushleft}

\end{document}